\documentclass[9pt,shortpaper,twoside,web]{ieeecolor}
\usepackage{generic}
\usepackage{cite}
\usepackage{amsmath,amssymb,amsfonts}
\usepackage{algorithmic}
\usepackage{graphicx}
\usepackage{textcomp}
\usepackage{epstopdf}
\usepackage{bm}

\usepackage{color,soul}
\def\BibTeX{{\rm B\kern-.05em{\sc i\kern-.025em b}\kern-.08em
    T\kern-.1667em\lower.7ex\hbox{E}\kern-.125emX}}
\begin{document}
\title{Practical Explicit-time Stabilization of a Proportional Control System}
\author{Wen Yan (https://orcid.org/0000-0002-0783-7837) and Tao Zhao.
\thanks{This study was first presented in a work report of the laboratory on November 3, 2023
\newline Submitted on: xx/xx/2024. This work was supported by Sichuan Science and Technology Program under Grant 24NSFSC0239. (Corresponding author: Tao Zhao.)}
\thanks{Wen Yan and Tao Zhao are with the College of Electrical Engineering, Sichuan University, Chengdu 610065, China (e-mail: yanwenessay@stu.scu.edu.cn (yanwenessay@126.com), zhaotaozhaogang@126.com). }
}

\maketitle

\begin{abstract}
Proportional control can be realized directly through the amplification of analog signals, and it also has the advantage of easy tuning parameters in digital signal control. However, it is difficult for the proportional control to preset the upper bound of settling time. To address this problem, a novel practical explicit-time control method is proposed. In bounded initial condition, this method makes this system error converge to a predefined neighborhood of zero within an explicit time. More specifically, the initial condition set and conditionally stable set are solved by practical explicit-time stabilization theorem. Based on that, a proportional feedback control is founded to achieve practical conditional fixed-time stability.
\end{abstract}


\section{Introduction}

Proportional Integral Differential (PID) control had the characteristics of simplicity, reliability and easy to implement, where the proportional control was often used to adjust the accuracy and response speed of the controlled system. In the case of strong radiation, strong interference and high signal delay, etc., control engineering was mainly based on analog circuit elements. Fortunately, proportional control can be achieved directly through amplifiers in analog circuits. In digital control, proportional control also had the advantage of easy parameter adjustment. However, traditional proportional control was generally based on experience, so that it was difficult to predefine the settling time.

Motivated by the above-mentioned issue, a novel explicit-time proportional control method is proposed to obtain better engineering advantage in control input. Meanwhile, the theoretical upper bound of settling time does not act as a cost. The main innovation is that, a direct proportion function is found to construct the practical conditional fixed-time stable system, which can stabilize a smooth proportional control system to predefined accuracy within explicit time.

\section{Preliminaries}

\begin{figure}[!t]
\centerline{\includegraphics[width=8.5cm]{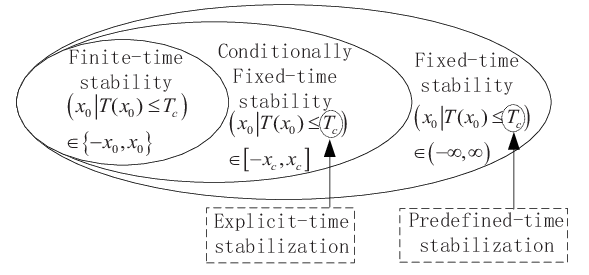}}
\caption{Block diagram of four stability theories on the relationship between settling time and initial state. $T_c$ represents the upper bound of settling time.}
\label{fig1}
\end{figure}

{\bf Definition 1} \cite{bhat2000finite,polyakov2014stability}: Consider a system:
\begin{equation} \dot x(t) = f(x(t)), x(0) = x_0.\label{eq:1}\end{equation}
where $f(x(t)):D \to R$ is continuous on an open neighbourhood $D$ of origin, and $f(0)=0$. System \eqref{eq:1} can be said to be finite-time stable if it is Lyapunov stable and finite-time converging from $D$. i.e. $\mathop {\lim }\limits_{t \to T(\bm x_0)} x(t) = 0$ and $T( x_0)$ is the settling-time function.

{\bf Definition 2} \cite{polyakov2011nonlinear}: If the system \eqref{eq:1} is finite-time stable and the settling time is bounded, then it can be called fixed-time stable. i.e. $\exists T_{max}>0: T( x_0)\le T_{max}, \forall x_0 \in D$.

{\bf Definition 3} \cite{kazda1962theoretical,kuo1995automatic}: If the stability of system changes as the closed-loop gain changes, the system can be called conditionally stable.

{\bf Definition 4}: If the system \eqref{eq:1} is finite-time stable and the settling time has a boundary that varies with the boundary of initial condition, then it can be called conditionally fixed-time stable. i.e. $\exists 0<T_c<+\infty: T( x_0)\le T_c, \forall x_0 \in D_c \subseteq D \backslash \{\infty\}$.

{\bf Remark 1}: Similar to conditional stability, the conditional parameters have effects on the settling time of conditionally fixed-time stable system. The fixed-time conditional parameter affects the boundary of the settling time, and the conditional stability parameter affects the stability of the system. \hfill $\square$

{\bf Lemma 2} \cite{yan2023explicit}: If a control system is conditionally fixed-time stable with explicit settling time function, then the design of control can be called explicit-time stabilization.

{\bf Remark 2}: The difference of the definitions between finite-time stability, conditionally fixed-time stability and fixed-time stability is that: the initial conditions that ensure the same prior upper bound for the settling time are different. It can be shown by Fig. \ref{fig1}.  \hfill  $\square$


{\bf Condition 1}: For most physical control systems, there is an upper bound on the states and its any-order derivatives. $x_0 \in D_c =[-x_c, x_c]$, and $x_c \in R^+$.

{\bf Lemma 3}\cite{yan2023explicit} : A class of conditionally fixed-time stable system with generalized explicit settling time function can be summarized as:
\begin{equation}{\dot  x} =  - \frac{G(V_e(x_c)^m) - G(0)}{T_c}{\left[\frac{{\partial G(V_e(x)^m)}}{\partial x}\right]^{ - 1}}.\label{eq:2}\end{equation}
where $0< m < 1$, $T_c  > 0$. $x=x(t)\in R$ is the system state. $x_0 \in D_c =[-x_c, x_c]$ represents the initial condition boundary of Condition 1. $V_e(x): D\backslash \{0\} \to R^+$ is the positive definite function and $V_e(0)=0$. System \eqref{eq:2} is conditionally fixed-time stable if the conditions a),b) hold:

(a): $G(V_e(x)^m)$ is continuous on $x\in R$.

(b): $\left[ {\frac{\partial G(V_e(x)^m)}{\partial V_e(x)^m}} \right]^{ - 1}$ is positive, bounded and right-continuous on $x\in D_c$.

Proof: Consider a candidate Lyapunov function as $V_e=\frac{1}{2}x^2$. The time derivative of $V_e$ can be expressed as:
\begin{equation}\begin{array}{l}
\dot V_e=x \dot x\\
~~~~=  x \cdot \left\{ - \frac{G(V_e(x_c)^m) - G(0)}{T_c}{\left[\frac{{\partial G(V_e(x)^m)}}{\partial x}\right]^{ - 1}}\right\}\\
~~~~=  - \frac{{G(V_e(x_c)^m) - G(0)}}{{{mT_c}}}{\left[ {\frac{{\partial G({V_e^m})}}{{\partial {V_e^m}}}} \right]^{ - 1}}V_e^{1-m}
\end{array}\label{eq:3}\end{equation}

Condition (b) points that $\left[ {\frac{\partial G(V_e(x)^m)}{\partial V_e(x)^m}} \right]^{ - 1}$ is bounded on $x\in D_c$. Hence, according to Definition 1 and \eqref{eq:3}, the system \eqref{eq:2} is finite-time stable. From conditions (a) and (b), we can deduce that $G(\cdot)$ is a monotone-increasing continuous function on $x\in D_c$. Then, the settling time can be deduced by \eqref{eq:3}:
\begin{equation}\begin{array}{*{20}{l}}
T(x_0)=\int_{V_e(x_0)}^{0} -{\frac{{m{T_c}}}{{[G(V_e(x_c)^m) - G(0)]{{\left[ {\frac{{\partial G({V_e^m})}}{{\partial {V_e^m}}}} \right]}^{ - 1}}{V_e^{1 - m}}}}} dV_e \\
~~~~~~~~=T_c \frac{G(\frac{1}{2^m}x_0^{2m})-G(0)}{G(\frac{1}{2^m}x_c^{2m}) - G(0)} \\
~~~~~~~~\le T_c
\end{array}\label{eq:4}\end{equation}

According to \eqref{eq:4}, the upper bound of the settling time can be conditionally preset as an explicit parameter, which is dependent of the initial condition boundary $\left|x_0\right| \le x_c$.

The proof is completed. \hfill $\blacksquare$

\section{Main result}

\subsection{The practical explicit-time stabilization of a state-constrained proportional control system}

\begin{figure}[!t]
\centerline{\includegraphics[width=5.5cm]{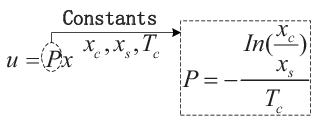}}
\caption{The diagram of proportional control.}
\label{fig2}
\end{figure}

Most of the existing finite-time and fixed-time control methods are based on the terminal attractor \cite{polyakov2015finite}. However, compared with the proportional control strategy, the smoothness and simplicity of those existing control methods are difficult to be guaranteed. Hence, inspired by the conditional stability theorem in Definition 3, a special conditionally fixed-time stable system structure is found to achieve explicit-time proportional control. Its principle is shown in Fig. \ref{fig2}.

Firstly, a practical conditionally fixed-time stable system can be summarized by Lemma 2 as:
\begin{equation}{\dot  x} = - \frac{G(V_{ep}(x_c)^m) - G(V_{ep}(x_s)^m)}{T_c}{\left[\frac{{\partial G(V_{ep}(x)^m)}}{\partial x}\right]^{ - 1}}.\label{eq:5}\end{equation}
where $x_s>0$ is the stable accuracy parameter such that $x_s<x_c$. $V_{ep}(x): D\backslash \{0\} \to R^+$ is the positive definite function and $V_{ep}(0)=0$. In this study, $V_{ep}(x)=\frac{1}{2 x^2_s}x^2$ holds.

Then, according to \eqref{eq:5}, a practical conditionally fixed-time stable system based on proportional gain can be written by selecting $G(V_{ep}(x)^m)= In(2^m V_{ep}(x)^m) =2m In( \frac{\left|x\right|}{x_s})$ as:
\begin{equation}\begin{array}{*{20}{l}}
\dot x = - \frac{In(\frac{x_c}{x_s})}{T_c}x
\end{array}\label{eq:6}\end{equation}

In the end, a smooth proportional control system can be designed to make system state reach the prescribed accuracy $\left| x \right|\le x_s$ within the explicit time $T_c$:
\begin{equation}\begin{array}{*{20}{l}}
\dot x = u_{ep}
\end{array}\label{eq:7}\end{equation}
in which
\begin{equation}\begin{array}{*{20}{l}}
u_{ep}= - \frac{In(\frac{x_c}{x_s})}{T_c}x
\end{array}\label{eq:8}\end{equation}

Proof:

Consider a candidate Lyapunov function as $V_{ep}=\frac{1}{2x_s^2}x^2$.

The time derivative of $V_{ep}$ can be expressed by \eqref{eq:7} and \eqref{eq:8} as:
\begin{equation}\begin{array}{l}
\dot V_{ep}=\frac{1}{x_s^2} x \dot x\\
~~~~~= \frac{1}{x_s^2} x \cdot \left[- \frac{In(\frac{x_c}{x_s})}{T_c}x\right]\\
~~~~~= - \frac{2In(\frac{x_c}{x_s})}{T_c}V_{ep}\\
~~~~~= - \frac{In(2^m V_{ep}(x_c)^m)}{mT_c}V_{ep}\\
\end{array}\label{eq:9}\end{equation}

Then, similar to Lemma 2, the reaching time of system state $x$ from domain $D_c$ to domain $D_s$ can be solved:
\begin{equation}\begin{array}{*{20}{l}}
T_{x_s}(x_0)=\int_{V_{ep}(x_0)}^{V_{ep}(x_s)} -{\frac{mT_c}{In(2^m V_{ep}(x_c)^m)}V_{ep}^{-1}} dV_{ep} \\
~~~~~~~~~~~=\int^{V_{ep}(x_0)}_{\frac{1}{2}} {\frac{mT_c}{In(2^m V_{ep}(x_c)^m)}V_{ep}^{-m}V_{ep}^{m-1}} dV_{ep} \\
~~~~~~~~~~~=\int^{V_{ep}(x_0)^m}_{\frac{1}{2^m}} {\frac{T_c}{In(2^m V_{ep}(x_c)^m)}V_{ep}^{-m}} dV_{ep}^m \\
~~~~~~~~~~~=T_c\frac{In(2^m V_{ep}(x_0)^m )-In(2^m \cdot \frac{1}{2^m})}{In(2^m V_{ep}(x_c)^m )}\\
~~~~~~~~~~~\le T_c
\end{array}\label{eq:10}\end{equation}
where $D_s=[-x_s,x_s]$ is a state space domain.

In the end, the following conclusions can be drawn:

1). According to \eqref{eq:9}, the system is global asymptotically stable.

2). According to \eqref{eq:10}, the upper bound of the practical stable time can be conditionally preset as an explicit parameter $T_c$ under a bound initial condition $D_c$.

The proof is completed. \hfill $\blacksquare$

\section{Theoretical analysis and simulation}

\subsection{A mathematical proof for the control input advantage of explicit-time control system over predefined-time control system}

The predefined-time stabilization control is popular one of many fixed-time stable control methods because of its parametric advantages. However, there exists a conservative gain term in this control system, which leads to an additional input problem \cite{yan2023explicit}. To solve this problem, the explicit-time stabilization control is applied under Condition 1. The input advantage can be mathematically demonstrated as follows:

Proof: Consider a comparison group of simple control systems:
\begin{equation}\left\{ {\begin{array}{*{20}{l}}
\dot x=u_{predefine}\\
\dot x=u_{explicit}
\end{array}} \right.\label{eq:19}\end{equation}

According to the results in \cite{jimenez2018note,liu2021trajectory}, a generalized predefined-time controller can be designed as:
\begin{equation}\begin{array}{*{20}{l}}
{u_{predefine}} =  - \frac{{G(\infty ) - G(0)}}{{{T_c}}}{\left[ {\frac{{\partial G({{\left| x \right|}^m})}}{{\partial x}}} \right]^{ - 1}}
\end{array}\label{eq:20}\end{equation}

According to the results in \cite{yan2023explicit,10315183}, a generalized explicit-time controller can be designed as:
\begin{equation}\begin{array}{*{20}{l}}
{u_{explicit}} =  - \frac{{G({{\left| {{x_c}} \right|}^m}) - G(0)}}{{{T_c}}}{\left[ {\frac{{\partial G({{\left| x \right|}^m})}}{{\partial x}}} \right]^{ - 1}}
\end{array}\label{eq:21}\end{equation}

According to the results in \cite{liu2021trajectory,yan2023explicit}, the candidate function $G$ about settling time should be monotonically increasing for predefined-time and explicit-time controllers. Then, by considering \eqref{eq:19}, \eqref{eq:20} and \eqref{eq:21}, the following inequality holds:
\begin{equation}\begin{array}{*{20}{l}}
\left| {{u_{predefine}}({t_0})} \right| - \left| {{u_{explicit}}({t_0})} \right| \\
= \frac{{G(\infty ) - G(0)}}{{m{T_c}}}\left[ {\frac{{\partial G({{\left| x \right|}^m})}}{{\partial {{\left| x \right|}^m}}}} \right]_{x = {x_0}}^{ - 1}{\left| x_0 \right|^{1 - m}}\\
~~~- \frac{{G({{\left| {{x_c}} \right|}^m}) - G(0)}}{{m{T_c}}}\left[ {\frac{{\partial G({{\left| x \right|}^m})}}{{\partial {{\left| x \right|}^m}}}} \right]_{x = {x_0}}^{ - 1}{\left| x_0 \right|^{1 - m}}\\
= \frac{{G(\infty ) - G({{\left| {{x_c}} \right|}^m})}}{{{T_c}}}\left[ {\frac{{\partial G({{\left| x \right|}^m})}}{{\partial {{\left| x \right|}^m}}}} \right]_{x = {x_0}}^{ - 1}{\left| x_0 \right|^{1 - m}} \\
\ge 0
\end{array}\label{eq:22}\end{equation}

In the end, according to \eqref{eq:22}, the initial control input of this generalized predefined-time controller is larger than this generalized explicit-time controller. The candidate function $G$ should be the same.

The proof is completed. \hfill $\blacksquare$

Relevant examples of simulation and experiment can be found in some existing studies \cite{yan2023explicit,10315183}.

\subsection{Simulation Validation}

Simulation is carried out in MATLAB2020A with fixed step and auto solver.

A system can be selected as:
\begin{equation}\begin{array}{*{20}{l}}
\dot x = x+u
\end{array}\label{eq:23}\end{equation}

A proportional controller can be designed as:
\begin{equation}\begin{array}{*{20}{l}}
u=- (\frac{In(\frac{x_c}{x_s})}{T_c}+1)x
\end{array}\label{eq:24}\end{equation}

By submitting \eqref{eq:24} to \eqref{eq:23}, with carry simulation in MATLAB2020A. Let $x_c=100$, $x_s=0.1$ and $T_c=1$. The performance curve of the system is shown in Fig. 3. According to the Fig. 3, the convergence accuracy of the system is less than 0.1, and the settling time is less than 1s.

\begin{figure}[!t]
\centerline{\includegraphics[width=7.5cm]{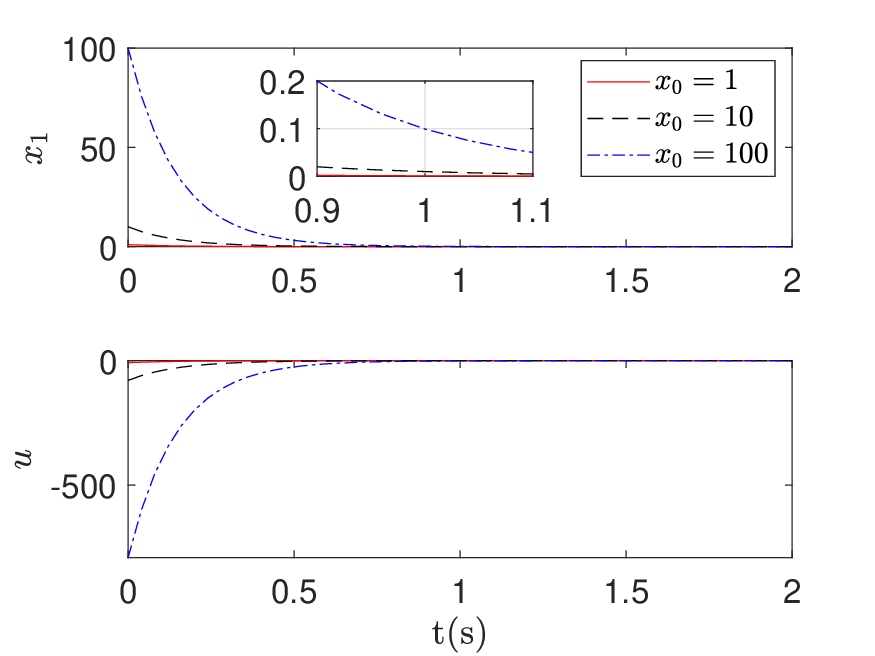}}
\caption{The state and control input curves with different initial condition.}
\label{fig3}
\end{figure}

\section{Conclusion}

A novel explicit-time proportional control method was proposed in this study. It can stabilize the system to a predefined accuracy within explicit time under bounded initial condition, which meant a practical conditional fixed-time stability. This control strategy was based on the idea of strict proportional control.

In the future, the proposed method will be applied in more practical projects to ease the conflict between input and settling time.

\bibliographystyle{IEEEtran}
\bibliography{IEEEexample}

\end{document}